\documentclass[12pt]{iopart}

\usepackage{iopams}  
\usepackage{graphicx}
\usepackage{hyperref}
\usepackage{lineno}
\usepackage{color}
\usepackage{url}

\begin{document}
\title{Method for environmental noise estimation via injection tests for ground-based gravitational wave detectors}
\author{T. Washimi$^{*,1}$, T. Yokozawa$^2$, T. Tanaka$^2$, Y. Itoh$^3$, J. Kume$^{4,5}$, J. Yokoyama$^{4,5}$}

\address{$^{1}$Gravitational Wave Science Project (GWSP), Kamioka branch, National Astronomical Observatory of Japan (NAOJ), Kamioka-cho, Hida City, Gifu 506-1205, Japan}
\address{$^2$Institute for Cosmic Ray Research (ICRR), KAGRA Observatory, The University of Tokyo, Kamioka-cho, Hida City, Gifu 506-1205, Japan}
\address{$^3$Osaka city university, Sugimoto Sumiyoshi-ku, Osaka City, Osaka 558-8585, Japan}
\address{$^4$Research Center for the Early Universe (RESCEU),  Graduate School of Science, The University of Tokyo, Hongo, Bunkyo-ku, Tokyo 113-0033, Japan}
\address{$^5$Department of Physics, Graduate School of Science, The University of Tokyo, Hongo, Bunkyo-ku, Tokyo 113-0033, Japan}
\ead{tatsuki.washimi@nao.ac.jp}

\begin{abstract}
Environmental noise is one of the critical issues for the observation of gravitational waves, 
but is difficult to predict in advance. 
Therefore, to evaluate the adverse impact of environmental noise on the detector sensitivity, 
understanding the detector response to the environmental noise in actual setup is crucial, 
for both the observation and future upgrades.
In this paper, we introduce and verify a new method of the environmental noise injection test based on the post-observation commissioning of KAGRA.
This new method (response function model) includes the frequency conversion and nonlinearity of power, 
which are the effects that are not considered in the current model (coupling function model) used in LIGO and Virgo. 
We also confirmed the validity of our method by applying it to an environmental noise-enriched dataset 
and successfully reproducing them.

\end{abstract}

\section{Introduction}\label{sec:Introduction}

\begin{figure}[htbp] \centering
  \includegraphics[clip,width=11cm]{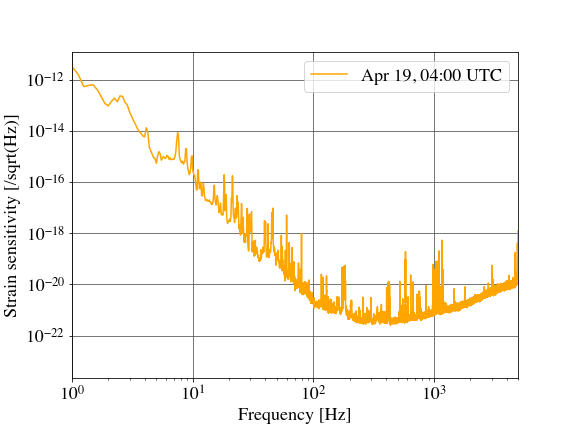}
  \caption{Typical amplitude spectral density (ASD) of the KAGRA strain sensitivity in O3GK~\cite{O3GK}.}
    \label{fig:Sensitivity_O3GK}
\end{figure}

Since the first detection of gravitational waves (GW) was achieved by the advanced LIGO~\cite{GW150914}, 
more than 50 GW events~\cite{GWTC-1,GWTC-2} have been detected by 3-detectors;  LLO, LHO in US, and Virgo in Italy. 
The 4th GW detector, KAGRA, constructed in Japan, is a unique detector that is in an underground facility 
and that cools the test-mass mirrors to reduce seismic noise and thermal noise~\cite{PTEP1}.
KAGRA performed the first joint observation run (O3GK) with GEO600 in Germany, from April 7 to 21, 2020~\cite{O3GK}. 
The typical strain sensitivity of the KAGRA interferometer in O3GK is shown in \Fref{fig:Sensitivity_O3GK}. 
Understanding the noise components, the so called "Noise budget", is important for distinguishing a GW signal from the noise 
and/or for improving the sensitivity of the detector.
This work is now ongoing, for example, in the context of the low-frequency region (approximately $< 100$~Hz) 
that is occupied by the control noise of the suspensions, 
whereas the floor level of the high frequency (approximately $> 400$~Hz) is consistent with the shot noise, 
and some peaks arise from the violin-modes of the suspension thermal noise or artificial lines used to control the interferometer.
The details will be published in the near future. 

A GW detector is exposed to many environmental noises, such as sounds, mechanical vibration, 
magnetic field, and sometimes they couple with the interferometer signal. 
For example, mechanical vibration of a vacuum chamber or an in-air optics induced by sound 
may contaminate the interferometer signal as a scattered light noise~\cite{scatter} or a beam-jitter noise.
Because the level of environmental noise depends on the time and the experimental site, and it is difficult to predict a priori, 
physical environmental monitoring (PEM) plays an important role in the observation of GW signals. 
To evaluate the environmental noise, we have installed many PEM sensors 
at the KAGRA experimental site (including outside the tunnel). 
Signals from the fast sensors (seismometers, accelerometers, microphones, magnetometers, and voltmeters with $\mathcal{O}(\mathrm{1~kHz})$ sampling, \textit{e.g.}, 2048~Hz sampling for the microphones) 
are acquired by the KAGRA digital system together with the interferometer signals and suspension signals.
The slow sensors (thermohygrometers and weather station with 1 minute sampling) have their own data loggers, 
and the signals are also merged into the KAGRA data recorded for every minute. 
The details of the KAGRA PEM before O3KG have been reported in an overview paper~\cite{PTEP3}.

\medskip

The power spectrum density (PSD) of the interferometer signal $S(f)$ can be divided into 
a component $S_\mathrm{PEM}(f)$ caused by environmental noise $P(f)$ monitored by a PEM sensor, 
and other noise $S_\mathrm{other}(f)$ that is independent of $P(f)$:
\begin{eqnarray}
S(f) = S_\mathrm{PEM}(f) + S_\mathrm{other}(f). 
\end{eqnarray}
One purpose of the PEM is to estimate the environmental noise $S_\mathrm{PEM}(f)$ for an observation mode.


\section{Concept of the PEM injection}\label{sec:Concept}

PEM injection is a technique to estimate the environmental noise $S_\mathrm{PEM}(f)$ 
in the interferometer signal $S(f)$ by increasing the environmental noise artificially.
In this paper, $S_\mathrm{bkg}(f)$ and $P_\mathrm{bkg}(f)$ denote PSDs for the strain signal 
and the PEM signal for the background data, respectively, 
and $S_\mathrm{inj}(f)$ and $P_\mathrm{inj}(f)$ denote those for the injection data. 

Because sometimes $S_\mathrm{inj}(f)$ is not larger than $S_\mathrm{bkg}(f)$ adequately, 
the transfer function is not available for PEM injection analysis. 
If $P_\mathrm{inj}(f)$ is sufficiently larger than $P_\mathrm{bkg}(f)$, 
$S_\mathrm{PEM}(f)$ can be derived, even though it is below the sensitivity $S_\mathrm{bkg}(f)$.

\subsection{The coupling function model}

A coupling function model has been developed in LIGO~\cite{LIGOpem, pemcoupling} 
and widely used in LIGO, Virgo~\cite{Virgo}, and KAGRA.
In this model, PEM projection $S_\mathrm{PEM}(f)$ for the background data is estimated as
\begin{eqnarray}
S_\mathrm{PEM}(f)
 = C^2(f) \cdot P_\mathrm{bkg}(f)  
 = \frac{ S_\mathrm{inj}(f) - S_\mathrm{bkg}(f) }{ P_\mathrm{inj}(f) - P_\mathrm{bkg}(f) }
   \cdot P_\mathrm{bkg}(f) , \label{eq:CouplingFunction}
\end{eqnarray}
where $C(f)$ is the coupling function
\footnote{For the ASD, $\sqrt{S_\mathrm{PEM}(f)} = C(f)  \sqrt{P_\mathrm{bkg}(f)}$}.
The excess in the interferometer $\Delta S = S_\mathrm{inj}(f) - S_\mathrm{bkg}(f)$ is not always significant.
In case of $\Delta S(f) < S_\mathrm{bkg}(f)$, the upper limit of the coupling function and 
the PEM projection are expressed as
\begin{eqnarray}
 C^2_{\mathrm{UL}}(f)
 = \frac{ S_\mathrm{bkg}(f) }{ P_\mathrm{inj}(f) - P_\mathrm{bkg}(f) }, \\
 S_{\mathrm{PEM,UL}}(f) =  C^2_{\mathrm{UL}}(f) \cdot P_\mathrm{bkg}(f) ,
\end{eqnarray}
instead of the coupling function and PEM projection themselves.
If the injected noise $\Delta P = P_\mathrm{inj}(f) - P_\mathrm{bkg}(f)$ is not sufficient, 
neither the PEM projection nor its upper limit are evaluated for such frequency.

This coupling function model is based on the following hypothesis: 
(1) Frequency conservation (no frequency conversion such as harmonics, side bands, etc.), 
(2) Linearity of PSD between the interferometer and environmental noise 
    (\textit{e.g.}, $S_\mathrm{PEM}$ doubles if $P(f)$ doubles), and
(3) Stability of the interferometer during measurement.
However, they are not always satisfied. For example, the scattered light noise is expressed as
\begin{eqnarray}
 h_{\mathrm{scat}}(t) = K \sin \left( \frac{8\pi}{\lambda} x(t) \right),
\end{eqnarray}
where $x(t)$ is the low frequency ($< 10$~Hz) vibration (displacement) of the surface on which the ghost beam is scattered (\textit{e.g.}, the inner surface of the vacuum chamber), 
$\lambda$ is the wavelength of light, and $K$ is a constant~\cite{scatter}. 
The PSD of $h_{\mathrm{scat}}(t)$ neither has linearity nor frequency conservation in general.
If there are some moving peaks or bumps that are independent of the environmental noise, 
they can be larger than the threshold of $\Delta S$ and make overestimation in the PEM projection result, it means some finite excess caused by instability (time dependence) of the interferometer itself and not be coming from the PEM injection. 


\subsection{The response function and non-linear model}

To include frequency conversion and nonlinearity, 
we expand the \Eref{eq:CouplingFunction} to the following formula:
\begin{eqnarray}
S_\mathrm{PEM}(f)
 = \int\! \Big[ R(f, f') \cdot P_\mathrm{bkg}(f') \cdot \varepsilon \Big] \mathrm{d} f' , 
\end{eqnarray}
where $R(f, f')$ is the response function
\footnote{The response function is widely used in the field of fast-neutron detector~\cite{ResponseFunction}.}, 
$\varepsilon=\varepsilon(f,f', P_\mathrm{bkg})$ is a function that describes some nonlinearity 
($\varepsilon=1$ for linear response). 

The coupling function model is included as
\begin{eqnarray}
 R_\mathrm{CF}(f, f') 
 = \frac{ S_\mathrm{inj}(f) - S_\mathrm{bkg}(f) }{ P_\mathrm{inj}(f') - P_\mathrm{bkg}(f') } 
   \cdot \delta(f-f'),
 \qquad \varepsilon=1.
\end{eqnarray}
When a single frequency $(f')$ environmental noise is injected, the kernel of the integral can be measured as follows:

\begin{eqnarray}
 R(f, f') \cdot \varepsilon 
 = \frac{ S_\mathrm{inj}(f) - S_\mathrm{bkg}(f) }{ P_\mathrm{inj}(f') - P_\mathrm{bkg}(f') } 
   \cdot \frac{1}{\Delta f'},
\end{eqnarray}
where $\Delta f'$ is the frequency resolution of these PSDs.
To take stability of the interferometer into account in the analysis quantitatively, 
the threshold of $\Delta S$ needs to be determined via a statistical treatment.

\section{Experimental setup}\label{sec:Setup}

\begin{figure}[htbp] \centering
  \includegraphics[clip,width=15cm]{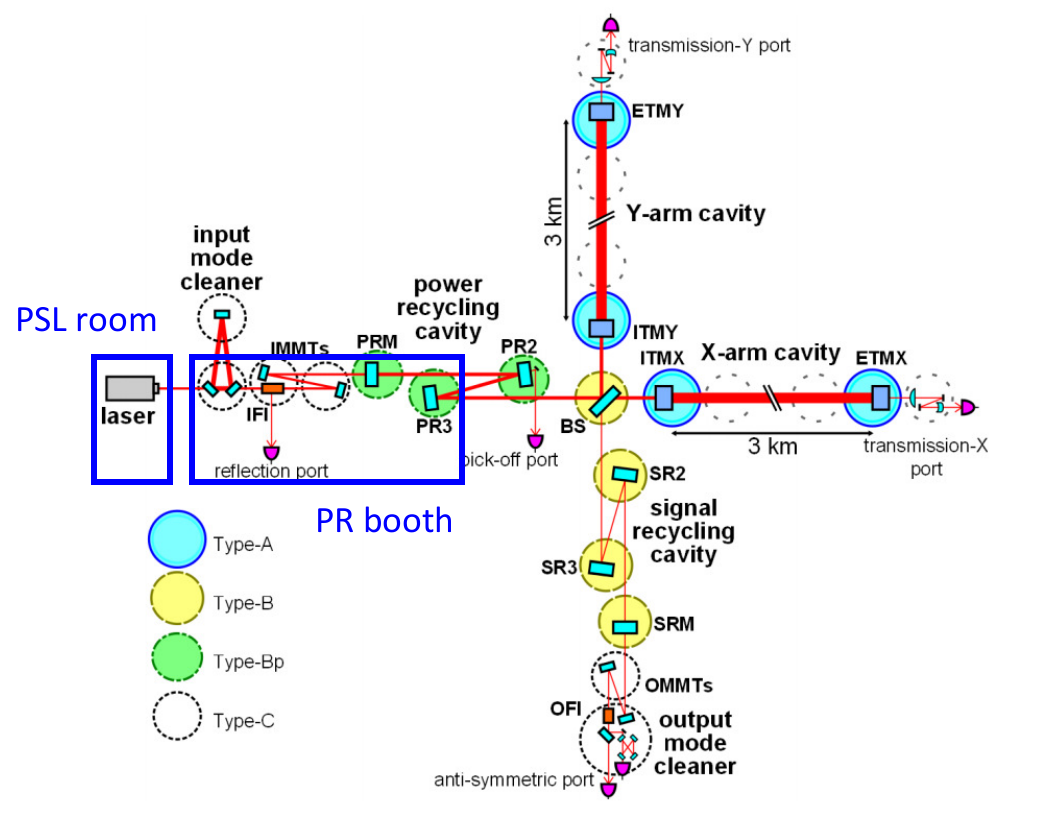}
  \caption{The names of clean booth in KAGRA, used in this study~\cite{PTEP1}.}
  \label{fig:CleanBooth}
\end{figure}

The measurements were performed on June 11, 2020, in the post-commissioning term of the O3GK. 
In this paper, the error signal (non-calibrated raw signal) of the KAGRA interferometer 
was used instead of the strain or differential arm length (DARM) because the accurate calibration 
was not realized for this day. 
\Fref{fig:CleanBooth} presents a schematic view of the KAGRA apparatus; 
laser path, mirrors, vacuum chambers, and clean booths.
In this study, we focused on the acoustic noise at the input optics ~\cite{IO} area, 
since it is known that the pre-mode cleaner in the PSL (pre-stabilized laser) room and 
the scattered light on the bellows between the IMC (input mode cleaner) and 
the IFI (input Faraday isolator) in the PR (power recycling) booth are sensitive to the acoustic noise, 
according to the experience of noise hunting in the pre-observation commissioning.
The microphones, speakers, and amplifiers used in this study are summarized in \Tref{tab:tools}.
\begin{table}\centering
\caption{\label{tab:tools}List of the PEM sensors and injectors used in this study.}
\begin{tabular}{llr}\br
    Description           & Brand name      & Operating frequency \\ \mr
    Microphone (PSL)      & B\&K 4188-A-021 & 20 -  12,500 Hz     \\
    Microphone (PR)       & ACO 7147A/4152  & 20 -  40,000 Hz     \\
    Microphone amp. (PSL) & B\&K 1704-A-002 & 22 -  22,400 Hz     \\
    Microphone amp. (PR)  & ACO TYPE5006/4  &  2 - 100,000 Hz     \\
    Speaker               & JBL JRX212      & 60 –  20,000 Hz     \\
    Speaker amp.          & QSC RMX5050a    & 20 –  20,000 Hz     \\ \br
\end{tabular}
\end{table}

\medskip 
The analysis in this study is based on gwpy 1.0.0~\cite{gwpy} and ROOT 6~\cite{ROOT}.
Most of the FFTs (except for \Fref{fig:Linearity}) were calculated using the Welch's method with a Hanning window, FFT length = 1~s, and overlap = 50\% 
\footnote{The Hanning window with 50\% overlap is the most general settings of FFT and we have no special reason to use other. FFT length = 1~s is chosen to balance between the frequency resolution and the number of averaging.}.
The PSDs of background are evaluated by approximately 5 min data.

\section{Frequency conversion and linearity} \label{sec:SingleLine}

\begin{figure}[htbp] \centering
  \includegraphics[clip,width=12cm]{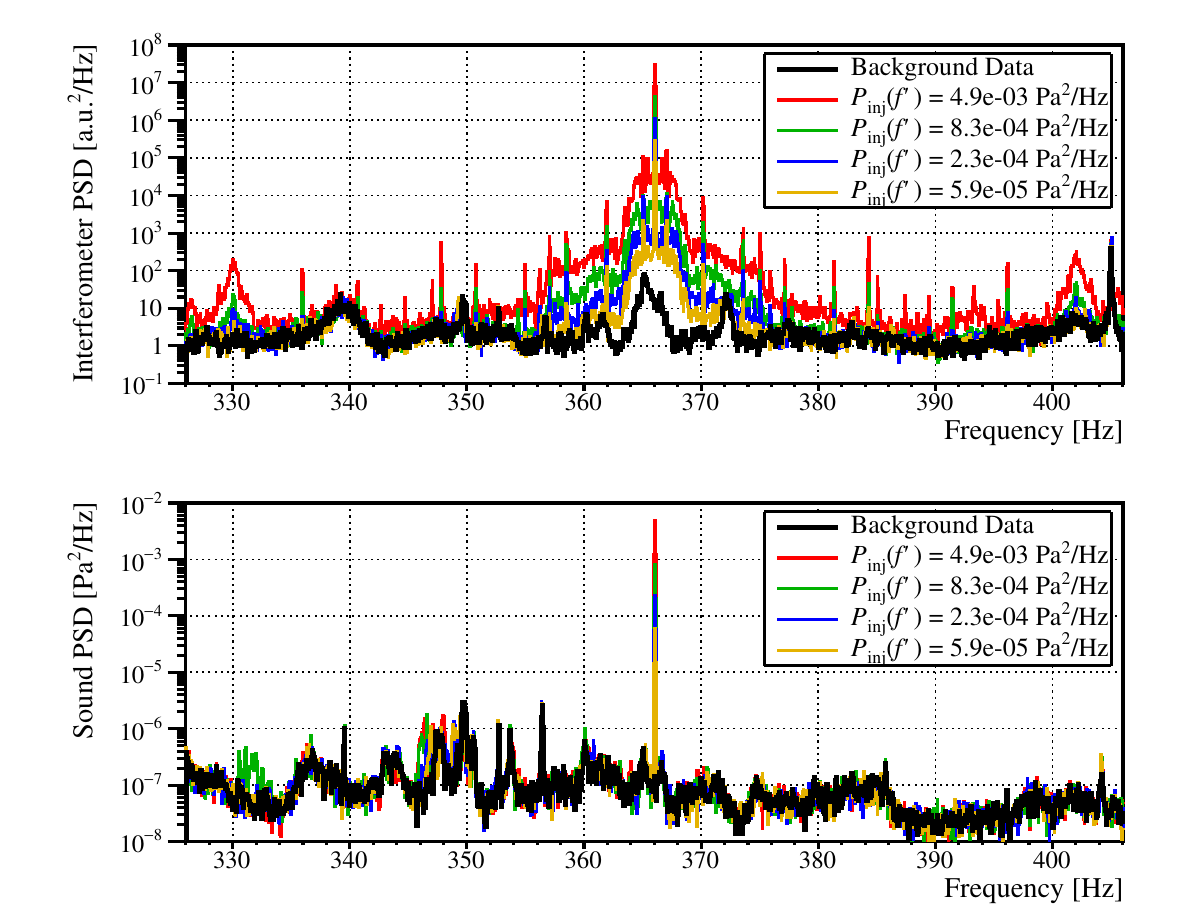}\\
  \includegraphics[clip,width=12cm]{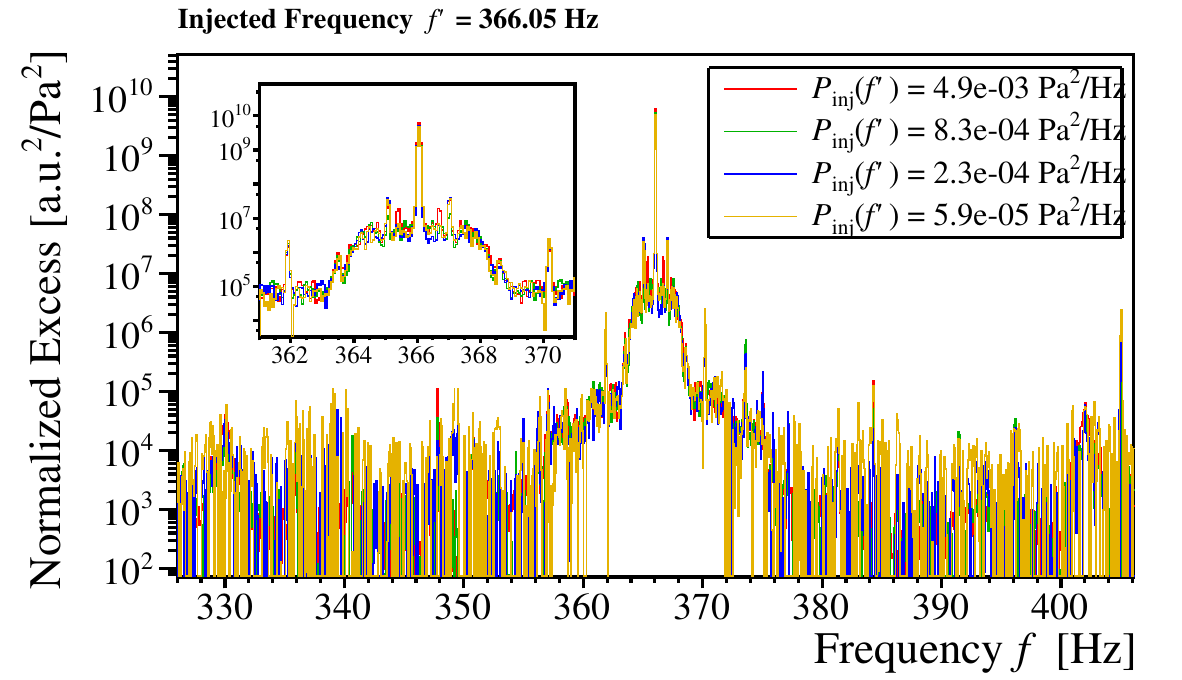}
  \caption{Top and middle : PSDs of the interferometer and microphone signal for the background (black) and 366~Hz single frequency acoustic injection (colors) data.  
  Bottom : Excess interferometer PSDs normalized by the microphone's signal at $f'=366~\mathrm{Hz}$ for each injection power. These PSDs were calculated with 16~s FFT and 10 average.}
    \label{fig:Linearity}
\end{figure}

Confirmation of the frequency conversion and PSD linearity are performed by single frequency 
PEM injection varying the injection power. 
\Fref{fig:Linearity} (top and middle) is an example where the result of the approximately 366~Hz 
single frequency acoustic injection in the PR booth is shown. 
The increase in the interferometer PSD around 366~Hz is due to the acoustic injection and is broad, 
even though the microphone signal has sharp frequency dependence. 
This result shows the evidence of frequency conversion for acoustic noise. 

The linearity between sound PSD and interferometer PSD can be verified by comparing 
the excess of the interferometer PSDs normalized by the microphone's signal:
\begin{eqnarray}
\left.  \frac{ S_\mathrm{inj}(f) - S_\mathrm{bkg}(f) }{ P_\mathrm{inj}(f') - P_\mathrm{bkg}(f') } \right| _{f'=366\mathrm{Hz}}
\end{eqnarray}
for various injection powers.
The result plotted in \Fref{fig:Linearity} (bottom) shows that this value is conserved,
at least up to $P(f'=366~\mathrm{Hz})=4.9\times 10^{-3}~\mathrm{Pa^2/Hz}$, 
which is approximately $10^5$ times larger than the background level. 
This means that the acoustic noise in the interferometer PSD $S_\mathrm{PEM}(f)$ is proportional to the sound PSD $P(f')$.
This behavior was also observed for other frequencies (harmonics, side bands, and other injection frequencies), 
and it is possible to set $\varepsilon=1$ for the following discussion.

\section{Evaluation of the response function and noise projection}\label{sec:Swept}

The value of the response function $R(f,f')$ can be evaluated by a single frequency ($f'$) PEM injection as follows:
\begin{eqnarray}
 R(f, f') 
 = \frac{ S_\mathrm{inj}(f) - S_\mathrm{bkg}(f)}{P_\mathrm{inj}(f') - P_\mathrm{bkg}(f') }
   \cdot \frac{1}{\Delta f'},
\end{eqnarray}
where $\Delta f'=1$~Hz is the bin width of $P(f')$.
Note that for the case of acoustic injection or radio wave injection, the signal detected by a PEM sensor
strongly depends on the location and frequency owing to the effect of multipath fading and/or shadowing. 
To cancel this effect, we use the approximated values of $P_\mathrm{inj}(f')$ 
instead of the measured value itself in this study. 
Details are described in \ref{sec:MIC_PSD}.

\begin{figure}[htbp] 
  \begin{minipage}{0.5\hsize} \centering
    \includegraphics[clip,width=8cm]{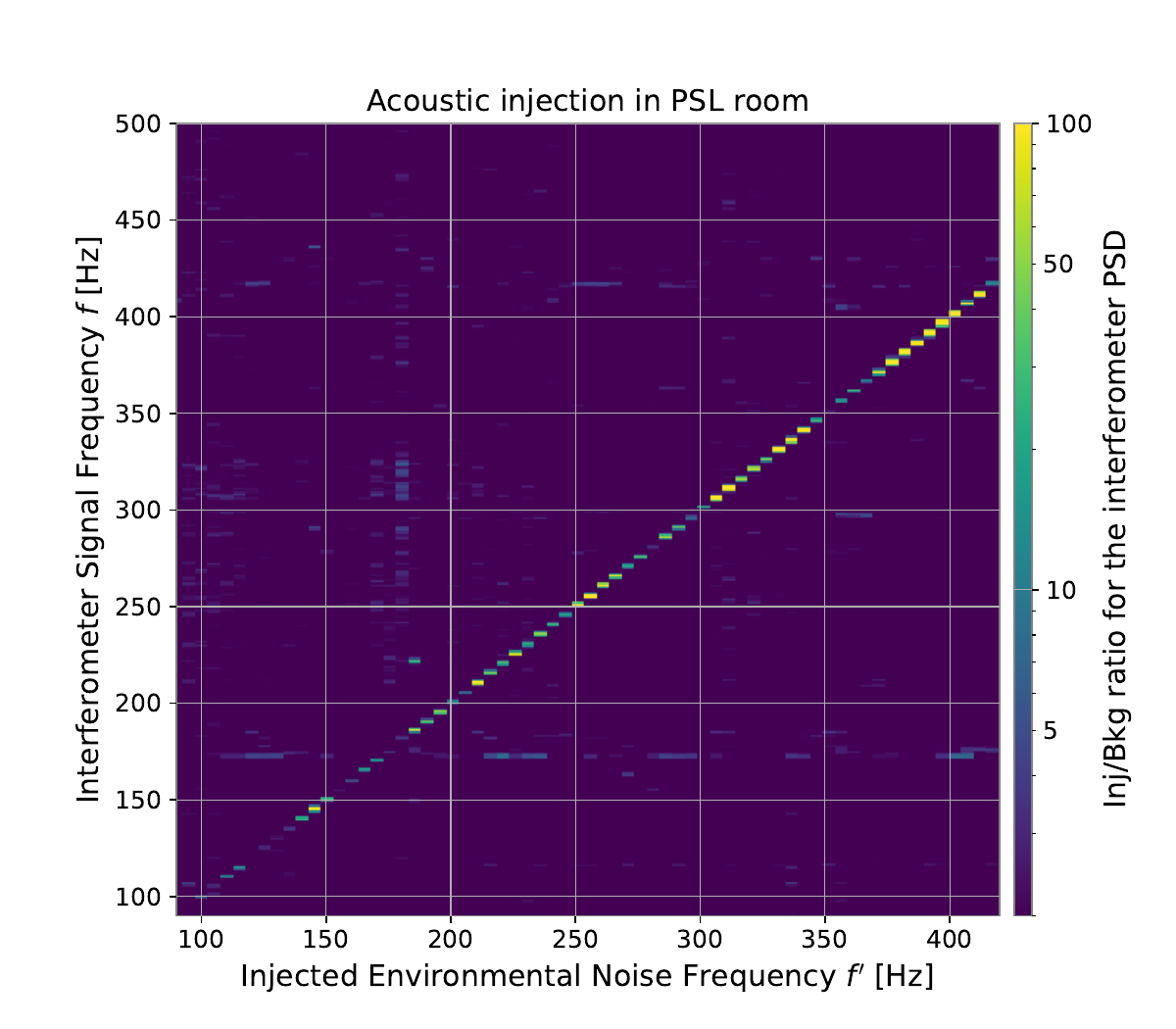}
  \end{minipage}
  \begin{minipage}{0.5\hsize}
    \includegraphics[clip,width=8cm]{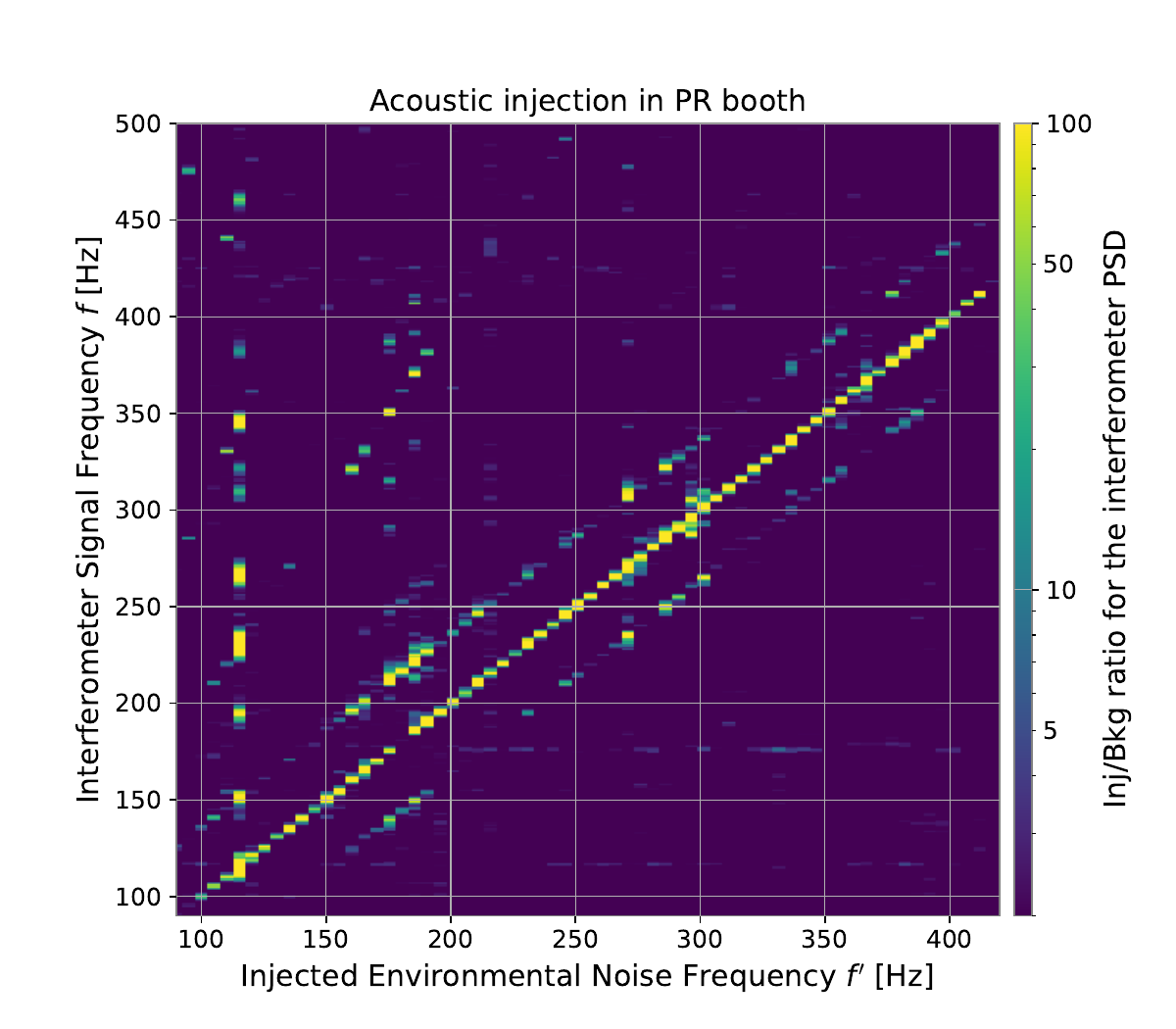}
  \end{minipage}
  \caption{Correlation between the injection frequency~$f'$ and output signal frequency~$f$ derived from the single frequency acoustic injections in the PSL room (left) and in the PR booth (right).}
    \label{fig:SNR_matrix}
\end{figure}

We performed single-frequency acoustic injections in the PSL room and PR booth,
with 200 frequencies from approximately 70.0~Hz to 1070.0~Hz, 10~s for each frequency, 
because the experimental time is not infinite. 
\Fref{fig:SNR_matrix} shows the correlation of the signal-to-noise ratio  $S_\mathrm{inj}(f) / S_\mathrm{bkg}(f)$ 
between the frequency of the injected noise and that of the interferometer signal. 
In the plot for the PSL room, a large excess was observed only in the line of $f=f'$, and there was little frequency conversion. 
However, in the plot for the PR booth, a clear excess was also observed on the side bands ($f=f'\pm36$~Hz), 
on the harmonics ($f=2f'$), or at a particular frequency ($f'=115$~Hz). 
This difference of the behaviour between the PSL room and the PR booth is
probably because in the former case, the sound shakes the optics (\textit{e.g.}, in-air mirror), 
whereas in case of the PR booth, the sound shakes the vacuum systems (\textit{e.g.}, chamber or bellows). 
Owing to the direct propagation of the vibrations to the interferometer in case of the PSL room, 
the interferometer signal is excited at the same frequency as the acoustic field.
By contrast, in case of the PR booth, frequency conversion occurs 
(i.e., the interferometer signal is excited at a frequency different from that of the acoustic field), 
owing to the indirect propagation of the vibrations to the
interferometer or because of other structures.

\begin{figure}[htbp] \centering
  \includegraphics[clip,width=16cm]{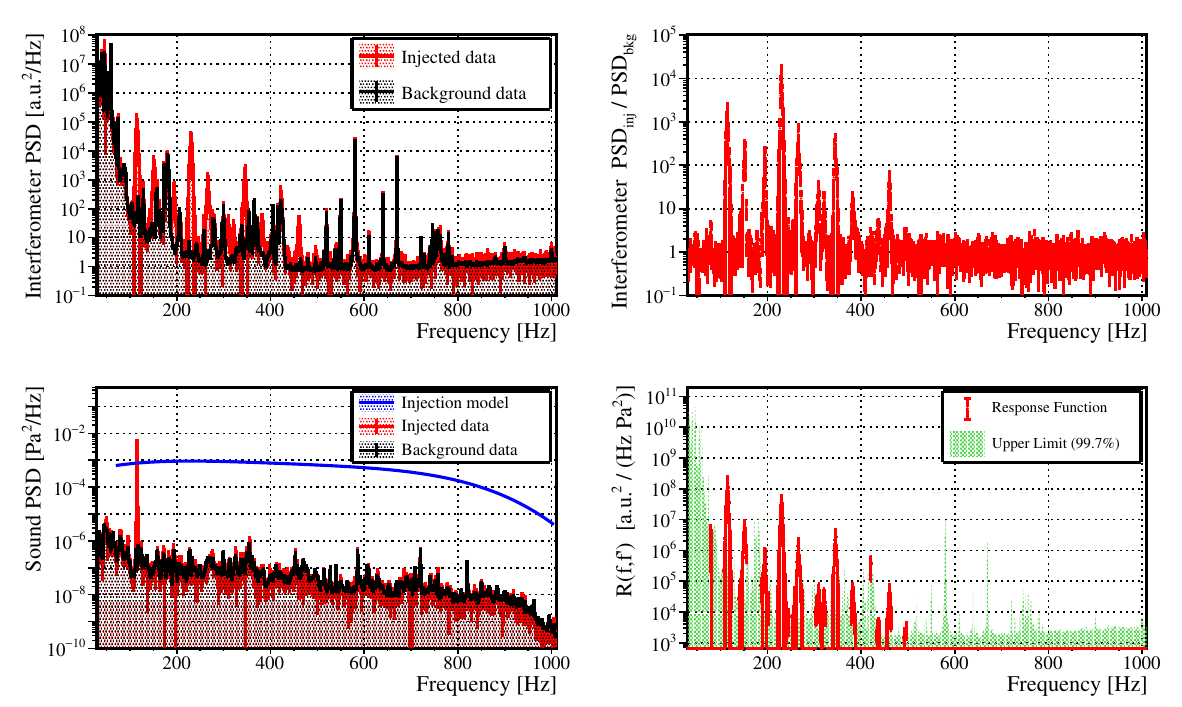}
  \caption{Snap shot of the single frequency acoustic injection in the PR booth at $f'=115$~Hz. 
  Top left : PSDs of the interferometer signal for injection data and background data. 
  Bottom left : Same as the microphone signal and the approximated function of the injected noise.
  Top right : Ratio of injection PSD and background PSD (SNR) for the interferometer signal. 
  Bottom right : Response function and its upper limit at $f'=115$~Hz.}
    \label{fig:PSDs_and_R_115Hz}
\end{figure}

When the difference $\Delta S(f) = S_\mathrm{inj}(f) - S_\mathrm{bkg}(f)$ 
is less than the threshold $\Delta S_\mathrm{th}(f)$,
the value of $R(f,f')$ is not defined for this frequency.  
Instead, the upper limits 
\begin{eqnarray}
 R_\mathrm{UL}(f, f') 
 = \frac{\Delta S_\mathrm{th}}{ P_\mathrm{inj}(f') - P_\mathrm{bkg}(f') } 
   \cdot \frac{1}{\Delta f'}
\end{eqnarray}
are calculated. 
The value of $\Delta S_\mathrm{th}(f)$ is defined statistically using the background data, 
as discussed in the \ref{sec:stat}. 
The error of $R(f,f')$ can be written as follows: 
\begin{eqnarray}
\delta R^2 &= \left[ \frac{1}{ P_\mathrm{inj} - P_\mathrm{bkg} } \cdot \frac{1}{\Delta f'} \right]^2
        \times \left[ \delta S_\mathrm{inj}\!^2 + \delta S_\mathrm{bkg}\!^2 \right] \\
&\qquad + \left[ \frac{R}{ P_\mathrm{inj} - P_\mathrm{bkg} } \right]^2 
          \times \left[ \delta P_\mathrm{inj}\!^2 + \delta P_\mathrm{bkg}\!^2 \right], \label{eq:dR^2}
\end{eqnarray}
where $\delta S_\mathrm{inj}$, $\delta S_\mathrm{bkg}$, $\delta P_\mathrm{inj}$, and $\delta P_\mathrm{bkg}$, 
are the statistical errors of each PSD and they are independent of each other. 
\Fref{fig:PSDs_and_R_115Hz} shows one snap shot of these calculations 
for the data with injected frequency $f'=115$~Hz. 
The injected acoustic noise monitored by the microphone had a PSD with a narrow peak without any harmonics, 
and was sufficiently larger than the background level ($P_\mathrm{inj} / P_\mathrm{bkg}\sim10^5$ at $f'=115$~Hz). 
Many peaks were observed in the PSD of the interferometer signal, not only at approximately 115~Hz, 
but also at the combination of the harmonics and sidebands. 
One remarkable point is that the SNR at $f=230$~Hz ( second harmonic) was larger than that at $f=115$~Hz.

The PEM projection $S_\mathrm{PEM}(f)$ and its upper limit are calculated as
\begin{eqnarray}
S_\mathrm{PEM}(f) 
 = \sum_{f'} \left[ R(f, f') \cdot \overline{P_\mathrm{bkg}}(f') \right]  \Delta f'_\mathrm{inj} , \\
\left[ S_\mathrm{PEM, UL}(f) \right]^2 = \sum_{f'} \left[ R_\mathrm{UL}(f, f') \cdot \overline{P_\mathrm{bkg}}(f') \cdot \Delta f'_\mathrm{inj} \right]^2 , \label{eq:UL_PSD}
\end{eqnarray}
where  $\Delta f'_\mathrm{inj} \sim 5$~Hz is the interval of injection frequency and
$\overline{P_\mathrm{bkg}}(f')$ is the average of $P_\mathrm{bkg}(f'')$ for 
$f'-\Delta f'_\mathrm{inj}/2 \leq f'' < f'+\Delta f'_\mathrm{inj}/2$.
The error of $S_\mathrm{PEM}(f)$ is a slightly complicated because 
$\delta R(f,f')$ also contains $n^2_\mathrm{bkg}$. 
However, since $P_\mathrm{inj}$ is much larger than $P_\mathrm{bkg}$ at the injected frequency, 
$P_\mathrm{bkg}$ and $\delta P_\mathrm{bkg}$ are negligible in the \Eref{eq:dR^2} : 
\begin{eqnarray}
\delta R^2 
\simeq \frac{ \delta S_\mathrm{inj}\!^2 + \delta S_\mathrm{bkg}\!^2}{(P_\mathrm{inj}\cdot\Delta f')^2} + \left[ \frac{R}{ P_\mathrm{inj} } \cdot\delta P_\mathrm{inj} \right]^2 .
\end{eqnarray}
Under this approximation, the error of the PEM projection can be written as
\begin{eqnarray}\fl
 \left[ \delta S_\mathrm{PEM}(f) \right]^2  
\simeq \sum_{f'} \left[ \left\{ \delta R(f, f') \cdot \overline{P_\mathrm{bkg}}(f')  \right\}^2 +  \left\{ R(f, f') \cdot \delta \overline{P_\mathrm{bkg}}(f') \right\}^2 \right]\cdot \Delta f'_\mathrm{inj} \! ^2 .
\end{eqnarray}
This error will be used in the discussion in the next section.

\begin{figure}[htbp] \centering
  \includegraphics[clip,width=10cm]{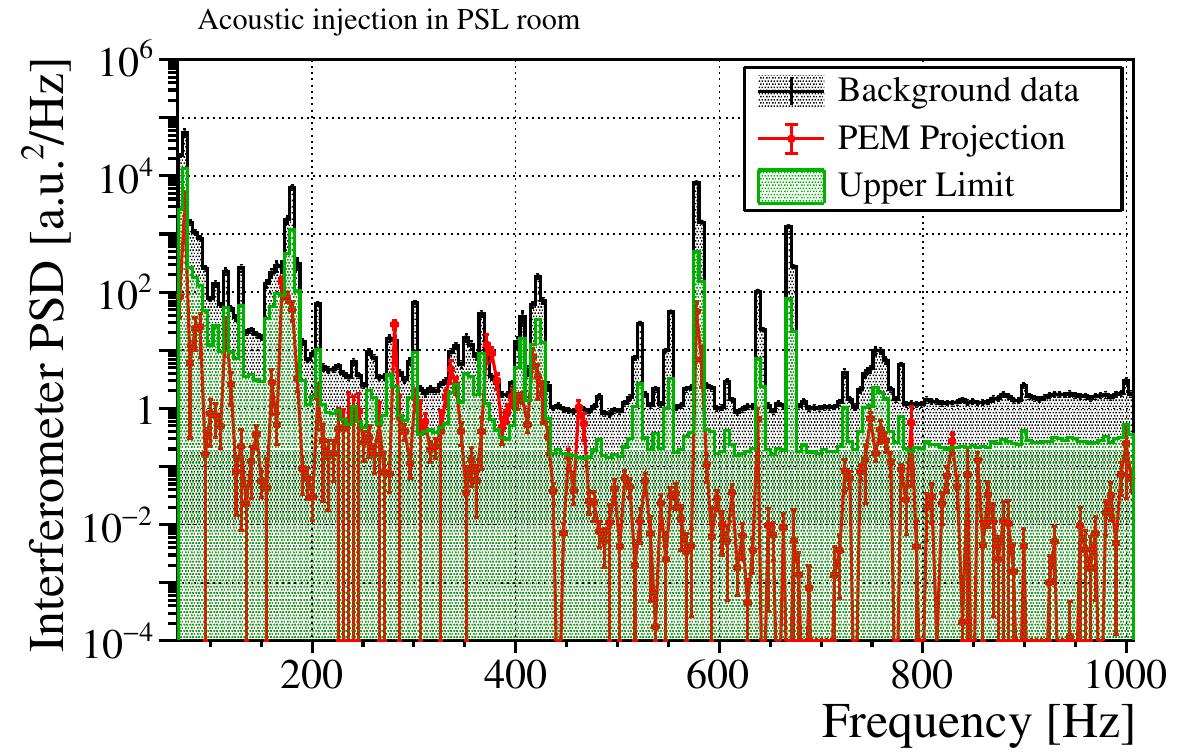}\\
  \includegraphics[clip,width=10cm]{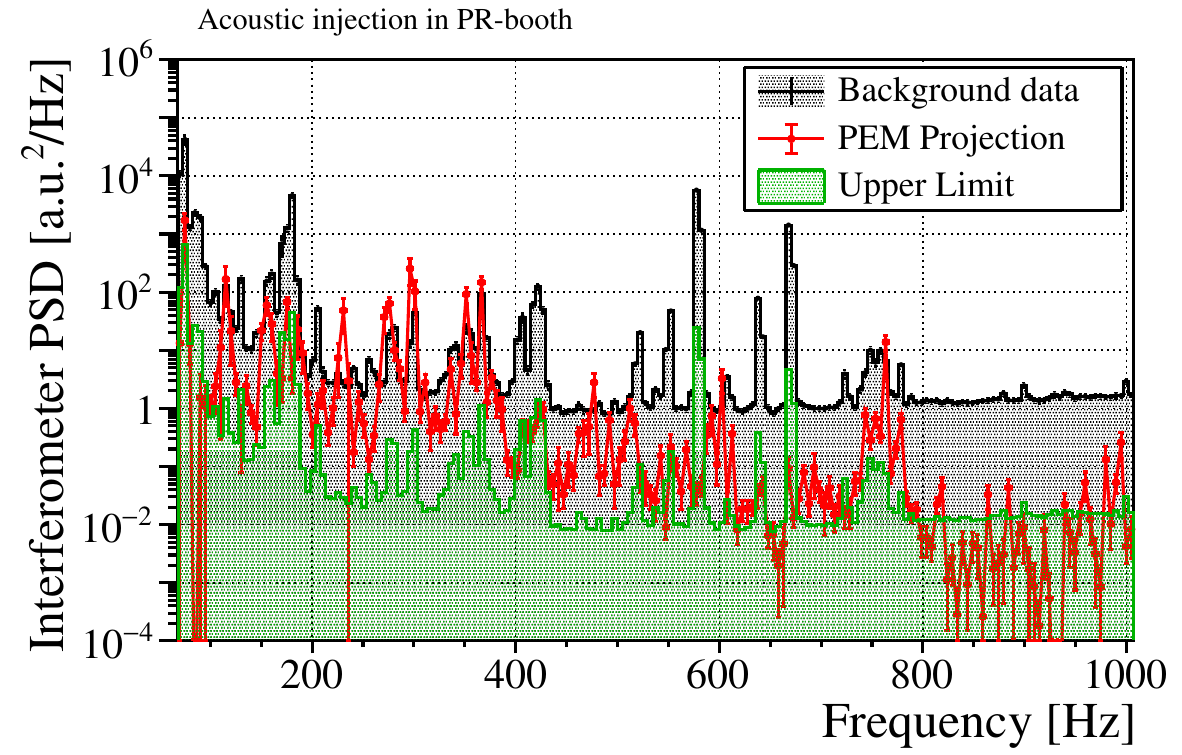}
  \caption{PEM projection of acoustic noise in the PSL room (top) and in the PR booth (bottom).}
    \label{fig:PEMprojection_bkg}
\end{figure}

\medskip
\Fref{fig:PEMprojection_bkg} shows the results of PEM projections 
and these upper limits of acoustic noise in the PR booth and in the PSL room.
On one hand, the projected acoustic noise in the PR booth was larger than the upper limit at most frequencies and was dominant around 200-400~Hz.
On the other hand, the projected acoustic noise in the PSL room contributed only around 350~Hz, and was smaller than the upper limit at most frequencies.

In case of the coupling function model, the situation is quite simple because we can calculate the PEM projection at the frequency for $\Delta S(f)>\Delta S_{\rm th}(f)$ and the upper limit for $\Delta S(f)<\Delta S_{\rm th}(f)$. 
However, in case of the coupling function model, the possibility that there are hidden excess below $\Delta S_{\rm th}(f)$ not only for the injected frequency ($f'$) but also for all frequencies ($f$) is considered. 
Finally, the upper limit is stacked up and became about $\times\sqrt{200}$ (the threshold $\Delta S_{\rm th}(f)$ is common for all $f'$) due to the integral of $f'$, where 200 is the number of injected frequencies. 
The results of PEM projection (red graph in \Fref{fig:PEMprojection_bkg}) is meaningful only they are above the upper limit (green).
Therefore, we need for this analysis (the response function model) to inject the noise with larger power compared to  the coupling function model case. 

We performed the acoustic injection with the same DAC count for both PSL room and PR booth. 
For the PSL room, constructed with hard door and wall, the speaker was located outside of the door to avoid defiling the cleanness.
But for the PR booth, we could locate the speaker inside of the clean booth. 
Actually, $P_\mathrm{inj}/P_\mathrm{bkg}$ was about$10^{4-5}$ for the PR booth but $P_\mathrm{inj}/P_\mathrm{bkg}$ was about $10^3$ for the PSL room. 
This is a good precept for us toward the further measurements.

\section{Test of the response function model by broadband acoustic injection}\label{sec:White}
\begin{figure}[htbp] \centering
    \includegraphics[clip,width=12cm]{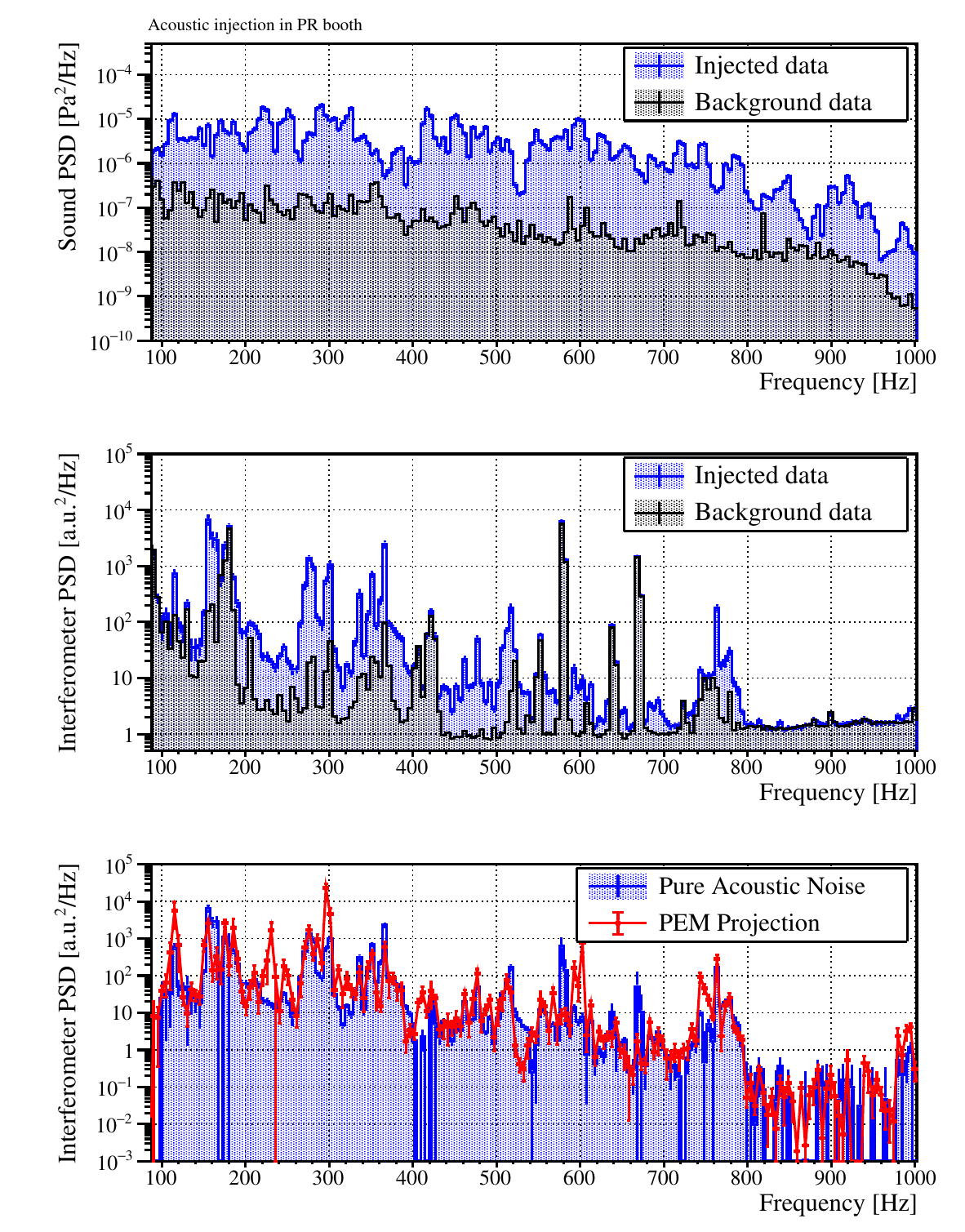}
  \caption{Results of the broadband acoustic noise injection test in the PR booth.
           Top: PSDs of the microphone signal. 
           Middle: PSDs of the interferometer signal.
           Bottom: The pure acoustic noise in the interferometer signal (blue) and the projection for them (red).}
    \label{fig:WhiteNoiseTest}
\end{figure}

Verification of the response function model and results of PEM projection were tested using another dataset. 
Here, $S_\mathrm{broad}(f)$ and $P_\mathrm{broad}(f)$ were the PSDs of 
the interferometer and the PEM sensor, respectively, for a broadband PEM injection. 
The difference $S_\mathrm{broad}(f) - S_\mathrm{bkg}(f)$ can be understood as a "pure environmental noise" in the interferometer signal when is is assumed that the other noises are stationary and canceled. 
This difference can be predicted as
\begin{eqnarray}
S_\mathrm{PEM}(f) 
 = \sum_{f'}  R(f, f') \left\{ \overline{P_\mathrm{broad}}(f') 
   - \overline{P_\mathrm{bkg}}(f') \right\}  \Delta f'_\mathrm{inj} , \label{eq:WhiteNoiseTest}
\end{eqnarray}
using the response function $R(f,f')$ derived from single-frequency injections. 
\Fref{fig:WhiteNoiseTest} shows the results of the broadband acoustic noise injection test 
in the PR booth, performed just after the single frequency acoustic injections. 
The excess in the interferometer signal is almost consistent with the calculation from the \Eref{eq:WhiteNoiseTest}.

However, there was some discrepancy larger than the statistical error in detail, \textit{e.g.}, around 230~Hz or around 600~Hz.
This can be interpreted as multipath fading and/or shadowing at the "weak points of the interferometer" (it means the points that the interferometer is affected by the sound effectively).
This situation will be improved by locating many microphones in the same area, or by performing acoustic injection from varied positions.

\section{Conclusion and future prospects}\label{sec:conclusion}

In this study, we performed the dedicated measurements concerning the response of 
KAGRA interferometer to the acoustic influence.
Using the single-line acoustic injection test, we clearly found that the linearity of the signal power 
was confirmed and that frequency conversion occurred. 
This means that as the coupling function model, the current model of environmental noise estimation
is not appropriate, and we, therefore, newly developed the response function model. 
We estimated the acoustic noise in the interferometer signal of KAGRA using our new method and confirmed its validity.
In the O3GK sensitivity, the acoustic noise in the PR booth was dominant around 200-400~Hz.

This method is quite general and can be applied not only for the acoustic injection, 
but also for other types of environmental noise injection.
We plan to perform it with more sensors and in more places in the next observation (O4).
This method can also be useful for other GW detectors as LIGO and Virgo.

To use the environmental information for an actual data analysis of GW search, 
it is necessary to estimate the time series of the environmental noise in the strain signal. 
Currently, we are working on it using independent component analysis (ICA), 
as we performed in iKAGRA data~\cite{ICA}.
Although the simplest linear mixing model has been investigated in this paper, 
ICA can be further extended to the case where the noise couples nonlinearly to the strain channel. 
This should be useful to deal with the acoustic noises observed in this work. 
We are going to improve ICA by appropriately taking into account the environmental information, 
and by establishing a noise subtraction scheme that enhances the efficiency of the GW search.

\section*{Acknowledgement}

This research has made use of data, software, and web tools obtained or developed by the KAGRA Collaboration.
In this study, we were supported by KAGRA collaborators, especially the commissioning members of the interferometer
and vibration-isolation systems, administrators of the digital system, and managers of the KAGRA experiment. 
We were also helped by the LIGO project, and the Virgo project, especially 
Robert Schofield \& Anamaria Effler in the LIGO PEM group 
and Federico Paoletti \& Irene Fiori in the Virgo environmental group. 
We would like to thank Editage (\url{www.editage.com}) for English language editing.

The KAGRA project is funded by the Ministry of Education, Culture, Sports, Science and Technology (MEXT) and 
the Japan Society for the Promotion of Science (JSPS).
Especially this work was founded by
JSPS Grant-in-Aid for Scientific Research (S) 17H06133 and 20H05639, 
JSPS Grant-in-Aid for JSPS Fellows 19J01299 
JSPS Grant-in-Aid for Scientific Research on Innovative Areas 6105 20H05256, 
and the Joint Research Program of the Institute for Cosmic Ray Research (ICRR) University of Tokyo 2019-F14, 2020-G12, and 2020-G21.

\appendix
\section{Smoothing the microphone's PSD}\label{sec:MIC_PSD}

As mentioned in \Sref{sec:Swept}, the apparent variation in the detected acoustic field is due to the frequency and locations of the speaker, the microphone, and other objects in the experimental area, even though the DAC count is constant, and the frequency characteristic of the speaker and the microphone are smooth.
This is caused by multipath fading and/or shadowing,  well known properties of the wave propagation. 
The number of paths of a wave from a source (the speaker in our case) to a detector (the microphone in our case) is not only one but many, due to the reflection and diffraction by the floor, walls, and other obstructions and they are interfered each other. 
The detected power is increased or decreased for some particular frequencies and it makes a complicated structure (depending on the positions of the wave source and the detectors are fixed) in the spectrum. This effect is multipath fading. 
Some obstructions larger than the wavelength ($\sim 1$~m for some hundred Hz sound) make shadows for the wave and the detected power is reduced behind of it. 
This is shadowing and its structure in the spectrum is larger than that of multipath fading. 

\begin{figure}[htbp] \centering
  \includegraphics[clip,width=16cm]{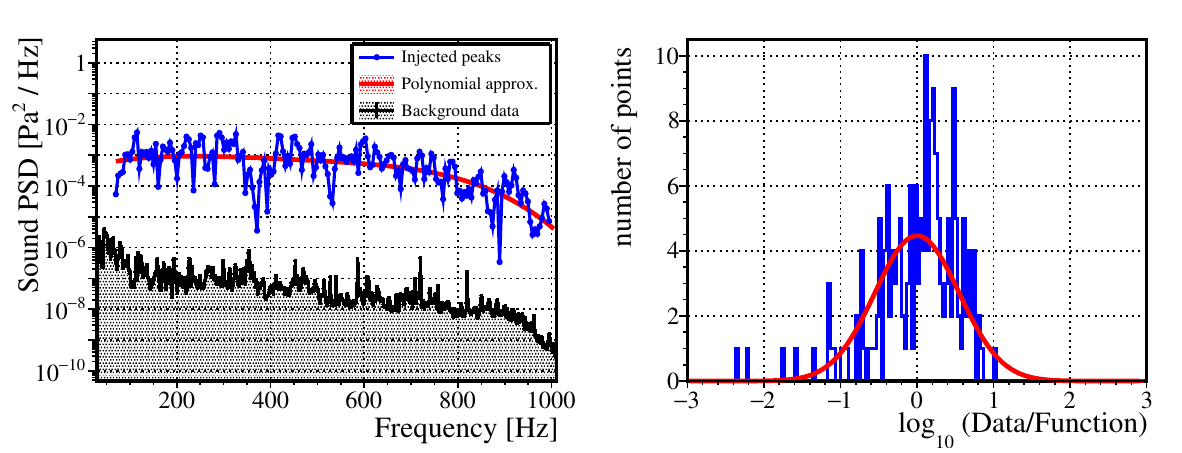}
  \caption{Left : PSD of the PR booth microphone for the background data (black), 
  the peak values for the single frequency acoustic injections (blue), 
  and the polynomial approximation (red).
  Right : Difference between the data and the function.}
    \label{fig:MIC_inj_func_PRbooth}
\end{figure}

\Fref{fig:MIC_inj_func_PRbooth} (left) shows the background spectrum of the microphone 
in the PR booth (black), and the peak values of each line for the single-frequency injection test (blue).
Because the points that the acoustic noise affects the interferometer are different from those of the microphone, this bias needs to be canceled.
Here, the data $\log_{10} P_\mathrm{inj}(f')$ is fitted to a polynomial function using the least-squares method. 
The red line in \Fref{fig:MIC_inj_func_PRbooth} (left) is the result, and 
the histogram in the \Fref{fig:MIC_inj_func_PRbooth} (right) is the difference between 
the data and the function.
The same procedure is also performed for the measurement in the PSL room.

\section{Statistical treatment}\label{sec:stat}
The threshold and the statistical error of the PSD value for each frequency are determined from the background data. 
Many "short span PSDs" $S_\mathrm{short}(f)$ are picked up from the background data randomly (approximately 10~s), 
instead of the injected data $S_\mathrm{inj}(f)$. 
The threshold $\Delta S_\mathrm{th}(f)$ is defined by the short span PSDs, 
at the 99.7\% percentile of the distribution for each frequency.

\begin{figure}[htbp] 
  \begin{minipage}{0.5\hsize} \centering
    \includegraphics[clip,width=8cm]{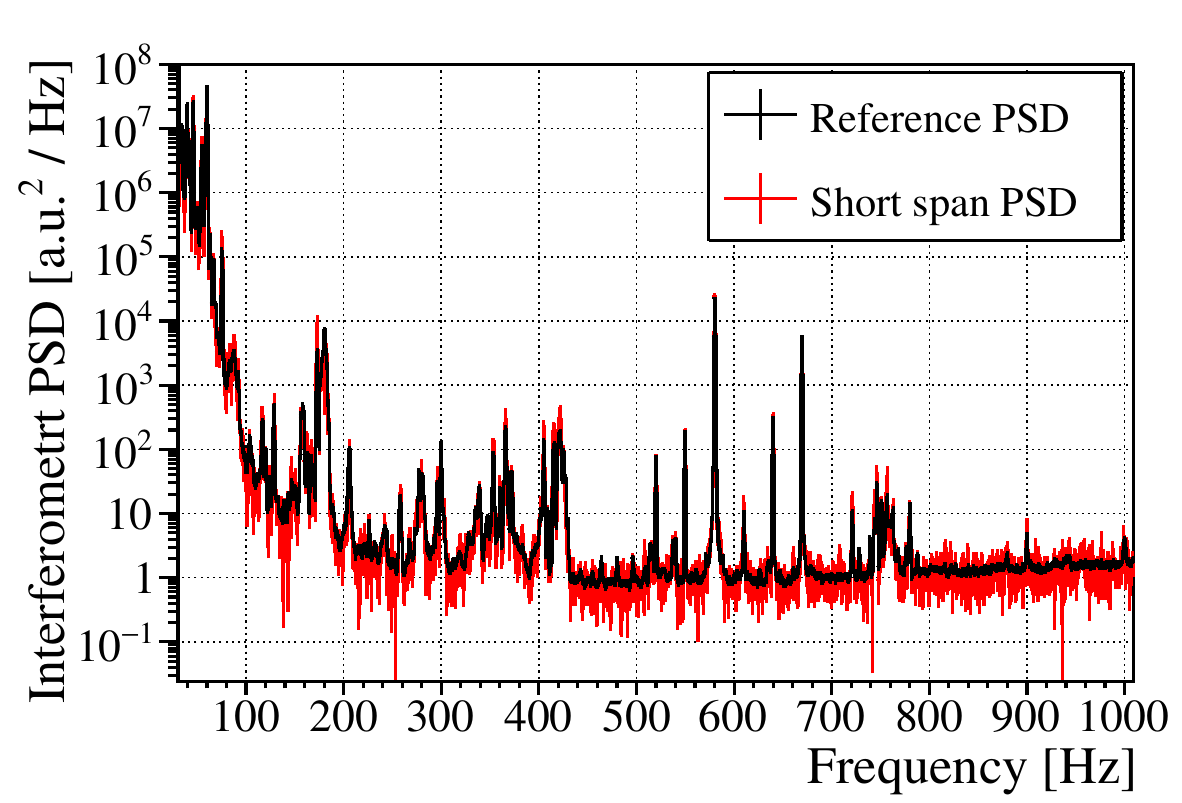}
  \end{minipage}
  \begin{minipage}{0.5\hsize} \centering
    \includegraphics[clip,width=8cm]{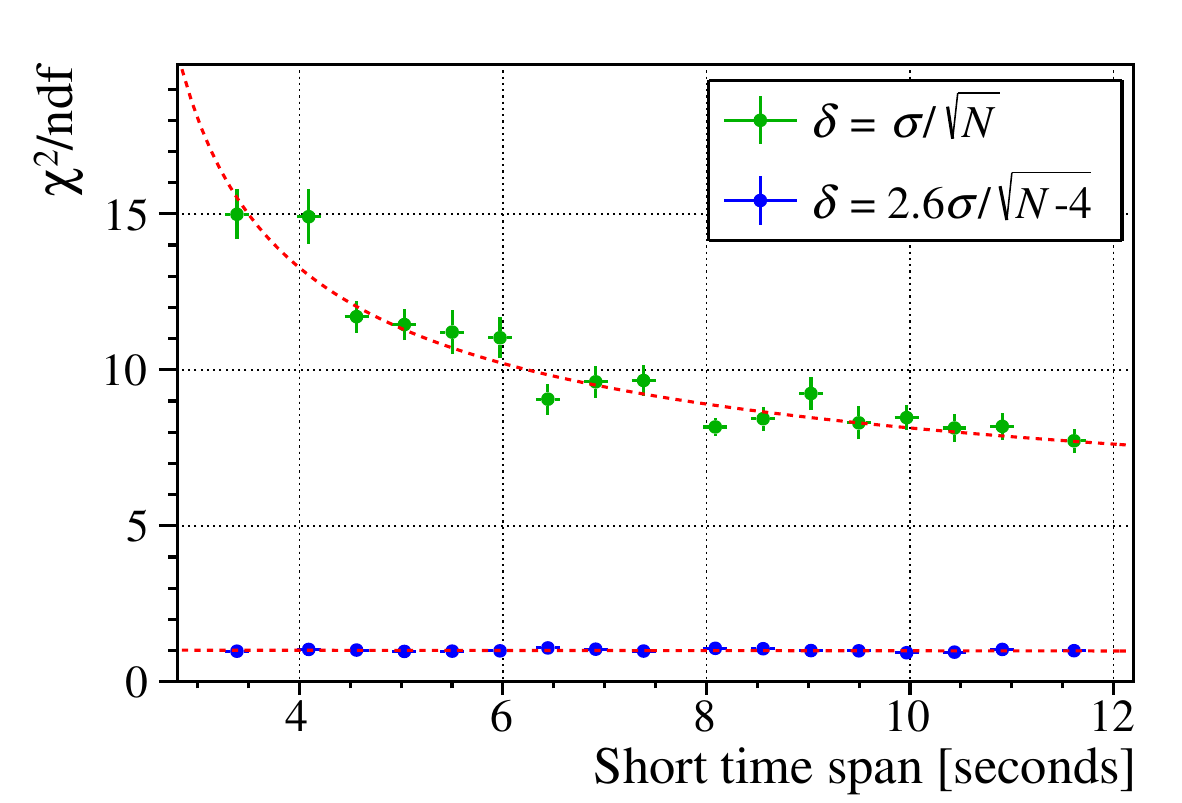}
  \end{minipage}
  \caption{Left : The reference PSD (black) and one short span PSD (red) of the interferometer signal for the background data.
  Right : Reduced $\chi^2$ between the reference PSD and short span PSDs as a function of the short time span, with $\sigma_S/\sqrt{N}$ error (green) and \Eref{eq:stat_error} error (blue). }
  \label{fig:ShortSpanPSD}
\end{figure}

\Fref{fig:ShortSpanPSD} (left) is one example of $S_\mathrm{short}(f)$ compared with the reference $S_\mathrm{bkg}(f)$. 
Since the PSDs are calculated by Welch's averaging method, the statistical error $\delta_{S}(f)$ 
for each frequency seems to be defined as $\delta_{S}(f) = \sigma_{S}(f)/\sqrt{N}$, 
where $\sigma_{S}(f)$ is the standard deviation, $N$ is the number of averages, 
if the signal is independent for each FFT window. 
We checked this definition via $\chi^2$ test. 
The green graph in \Fref{fig:ShortSpanPSD} (right) is the reduced $\chi^2$ between $S_\mathrm{short}(f)$ 
and $S_\mathrm{bkg}(f)$ with this error, averaged over a short time span $(\propto N)$. 
It should be unity for all $N$ when the error is defined appropriately, but it is not.
We modified the statistical error as
\begin{eqnarray}
\delta_{S}(f) = \frac{2.6\ \sigma_{S} (f)}{\sqrt{N-4}} , \label{eq:stat_error}
\end{eqnarray}
to make the reduced $\chi^2$ to be 1. 
The blue graph in \Fref{fig:ShortSpanPSD} (right) is the result.

\end{document}